\documentclass[runningheads]{llncs}
\usepackage{graphicx}
\usepackage{cite}
\usepackage{float}
\usepackage{amsmath,amssymb,amsfonts}
\usepackage{algorithmic}
\usepackage{graphicx}
\usepackage{tabularx}
\usepackage{textcomp}
\usepackage{xcolor}
\usepackage{graphicx}
\usepackage{multicol}
\usepackage{subcaption}
\emergencystretch=\maxdimen
\newcommand{\repeatthanks}{\textsuperscript{\thefootnote}}

\begin{document}
\title{DEEPFAKE CLI: Accelerated Deepfake Detection using FPGAs}
%
%
\author{Omkar Bhilare\thanks{These Authors contributed equally to this work}\inst{1,2}\and
Rahul Singh\repeatthanks\inst{1,2} \and
Vedant Paranjape\repeatthanks\inst{1,2} \and
Sravan Chittupalli\repeatthanks\inst{1,2} \and
Shraddha Suratkar\inst{1,2} \and
Faruk Kazi\inst{1,2}
}

\authorrunning{O. Bhilare, R. Singh, V. Paranjape et al.}
%
\addtocounter{footnote}{-1}

\institute{Department of Electrical Engineering, V.J.T.I, Mumbai, India. \and
\email{\{oabhilare\_b19,rsingh\_b18, vvparanjape\_b18\, schittupalli\_b18}@el.vjti.ac.in, sssuratkar@ce.vjti.ac.in, fskazi@el.vjti.ac.in}
\maketitle              
\begin{abstract}
Because of the availability of larger datasets and recent improvements in the generative model, more realistic Deepfake videos are being produced each day. People consume around one billion hours of video on social media platforms every day, and that’s why it is very important to stop the spread of fake videos as they can be damaging, dangerous, and malicious. There has been a significant improvement in the field of deepfake classification, but deepfake detection and inference have remained a difficult task. To solve this problem in this paper, we propose a novel DEEPFAKE C-L-I (Classification – Localization – Inference) in which we have explored the idea of accelerating Quantized Deepfake Detection Models using FPGAs due to their ability of maximum parallelism and energy efficiency compared to generalized GPUs. In this paper, we have used light MesoNet with EFF-YNet structure and accelerated it on VCK5000 FPGA, powered by state-of-the-art VC1902 Versal Architecture which uses AI, DSP, and Adaptable Engines for acceleration. We have benchmarked our inference speed with other state-of-the-art inference nodes, got 316.8 FPS on VCK5000 while maintaining 93\% Accuracy.

\end{abstract}
\section{Introduction}
Deepfake is artificially created media in which a frame is created synthetically using someone else's features like face, structure, lip movements ,etc. They are usually created by leveraging a Generative Adversarial Network to create a picture or video which looks realistic enough to deceive any person. 
While Deepfakes were initially created to prank individuals, they started getting attention due to their use in illegal activities like celebrity pornographic videos, fake news, and bullying. Hence, detecting deepfakes has become a very important issue in recent years. \\
Upon going through relevant literature and existing methods of deepfake classification, we observed that it is essentially an image classification problem, but the catch here is that pathological differences between the real and fake images are quite small, as a result existing CNN models need to be modified and tuned to detect these minute differences. \\
MesoNet is a CNN architecture which is used to detect Face2Face and Deepfakes manipulations accurately\cite{b3}. It is designed in such a way that it uses cropped faces from videos and analyses mid-level features. The model focuses on the right amount of details by using an architecture with small number of layers. XceptionNet is a complex model originally designed for working with 2D Images which uses depthwise separable convolutional layers with residual connections\cite{b4}, as a result it gives higher performance than MesoNet\cite{b3}. However, this model takes a lot of time to train. EfficientNets which are relatively a newer family of CNN models aimed at providing efficient resource management by balancing model parameters like width, depth, and resolution, outperform models which have a similar number of parameters\cite{b5}.
Researchers have taken a step further and proposed classifying each pixel of the image as real or fake. The U-Net architecture addresses this issue by employing an encoder-decoder based network with skip connections\cite{b6}. To improve classification accuracy, Eff-Ynet describes a novel architecture which combines EfficientNet encoder with a classification and segmentation branch\cite{b7}. It is designed to classify an image and also find regions where the image is real and where it is fake. The job of segmentation helps train the classifier, and at the same time it also produces useful segmentation masks. \\
The inference of neural networks is usually slow on general-purpose GPUs\cite{a1}. One way to accelerate the inference of these networks is to use soft cores emulating on the FPGAs, which allows the user to utilize the task and data level parallelism to reach the performance of ASIC implementations while taking reduced design time\cite{b9}. It is also possible to make RNN or CNN specific hardware architectures. In this paper, they have accelerated Deep Recurrent Neural Network (DRNN) on a hardware accelerator running on XILINX ZYNQ FPGA\cite{b10}. These ZYNQ boards are heterogeneous in nature, which means it has a hardcore CPU besides FPGA. In one of the paper, researchers have mapped BNNs (Binarized neural networks) onto an FPGA device while FP32 networks are mapped to the CPU, this hybrid mapping increases overall neural network efficiency while maintaining inference speed\cite{b11}. Researchers have found these ZYNQ boards perform better than CPUs and GPUs\cite{b12}.\\
One of the paper, transforms the model into FP8\footnotemark[1]\footnotetext[1]{FP8 means 8-bit floating point representation} format for speed up\cite{8bit}. Upon evaluating this, we decided to go with INT8\footnotemark[2]\footnotetext[2]{INT8 means 8-bit integer representation} in our implementation, thus quantized our model to INT8 precision. This means that our model, which was originally trained on FP32\footnotemark[3]\footnotetext[3]{FP32 means 32-bit floating point representation} precision, was converted to INT8\footnotemark[1] precision. By reducing the precision, the number of bits required to store the model parameters are reduced, which means it requires less amount of memory to store and reduces the number of clock cycles to transfer data between memory and the accelerator over the PCIe bus. \\
We propose U-YNet, a combined segmentation, and classification model which consists of UNet Encoder and Decoder responsible for producing a segmentation map to show the altered regions in the deepfake content and a classification branch present at the end of the UNet Encoder branch responsible for classifying if the media content is real or deepfake origin. The segmentation map is a novel way to identify the regions of a face that have been mangled to create the deepfake, giving an insight into the construction of a deepfake and paving the way for creating models to reverse a deepfake as well. This model runs on a Deep Learning Processing Unit(DPU)\footnotemark[6] present in the AMD-XILINX VCK 5000 Versal FPGA device. It consists of dedicated processing units like hardware accelerated convolution engine which enables convolution based model like UNet (and U-YNet) to run with higher inference speeds, enabling real-time classification of deepfake content.

\section{Motivation and Background}
\begin{figure}[htp]
\begin{center}
\includegraphics[scale=0.2]{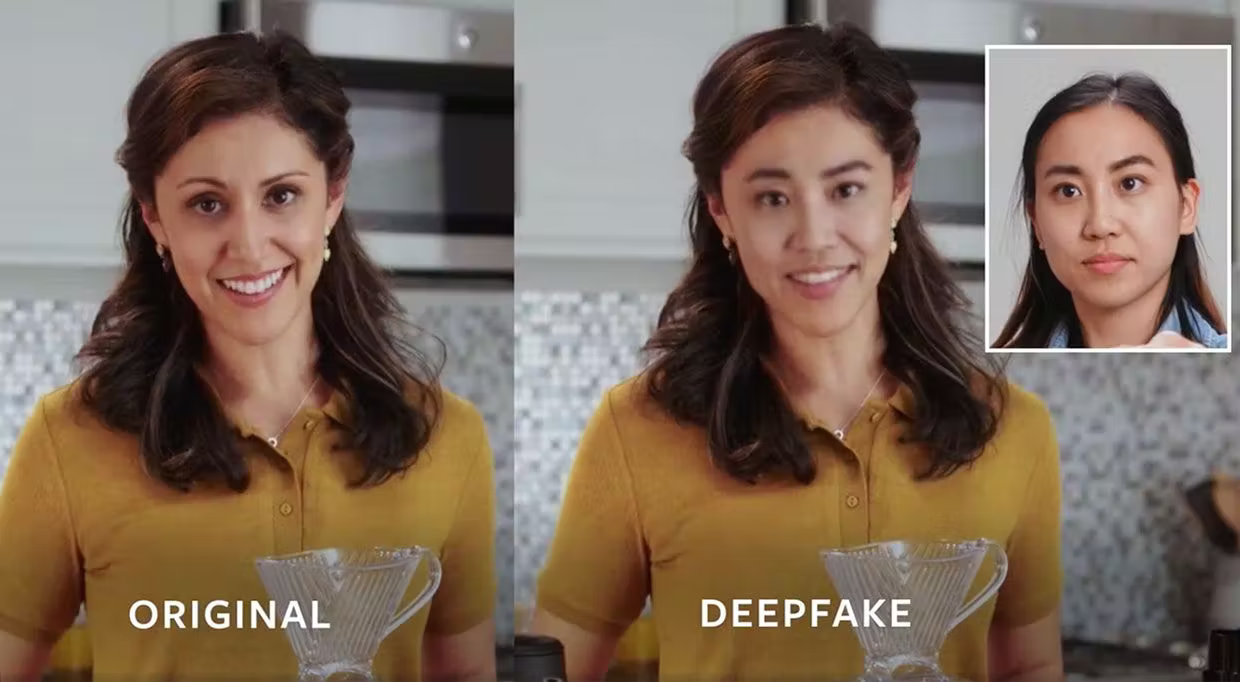}
\end{center}
\caption{Completely believable deepfakes can be generated with ease nowadays.}
\label{fig1}
\end{figure}

Improvement in generative models and abundance of datasets has led to evolution of models that can generate realistic looking deepfake videos, deceiving the human eye and machines as well (Fig.~\ref{fig1}). There is a huge potential to spread deepfaked videos by malicious actors for their gains, as more than 100 million hours of video content is watched every day on social media.

Looking at this with the perspective of computation, faster computation speeds and easily available compute resources means that deepfake videos can be generated with ease. This makes fast and effective detection of these videos very crucial to stop the malicious use of deepfake videos. \\
On March 16, 2022, a video claiming to show Ukraine President Volodymyr Zelensky calling for the Ukrainian people to surrender to Russia was aired on news channel Ukraine 24, and circulated on social media. Such events clearly show how potent of a tool, deepfakes are for creating social chaos. Thus, there is a need for faster detection of such deepfake videos so that they can be taken down even before reaching the social media users. This paper aims to leverage the massively parallel nature of FPGAs to efficiently run Deep Neural networks that power the deepfake detection algorithms.

\section{Implementation}
\begin{figure}
\begin{center}
\includegraphics[scale=0.35]{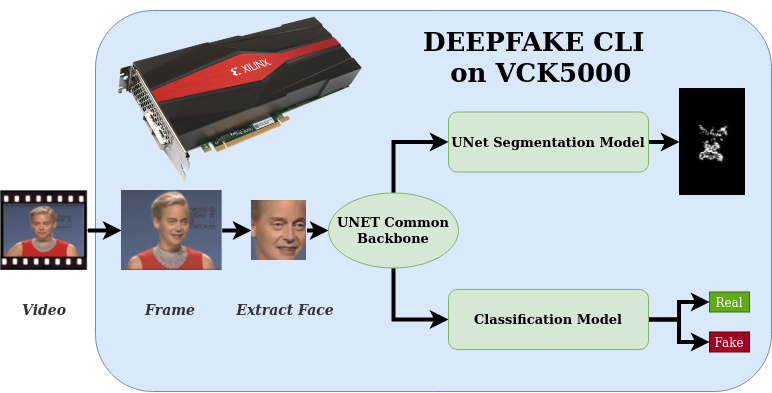}
\end{center}
\caption{Flowchart of Deepfake CLI}
\label{fig2}
\end{figure}

Deepfake C-L-I (Classification – Localization – Inference) is a novel system that we are proposing in this paper, which consists of an optimum combination of software and hardware tools to accelerate deepfake detection on FPGAs. The block diagram of Deepfake CLI is shown in Fig.~\ref{fig2}.
The software part of the acceleration system consists of a deep learning model comprised of a UNet Segmentation and a Classification Model. As for the hardware part, these models are first quantized to INT8 for DPU Architecture supported by VCK5000 to achieve maximum performance, then later on it is run on the FPGA. \\
The implementation of Deepfake C-L-I is divided into two parts as software and hardware as follows:

\subsection{Software Implementation}
We implemented a multitask model called U-YNet, that can produce classification (real or fake) and segmentation results on deepfake content. The backbone which is the UNet Encoder takes a frame as the input and then the encoded features are passed to the classification branch and the UNet decoder simultaneously. The reason for doing this is that the combination of loss of classification and segmentation tasks may help the classification branch to learn from the segmentation branch features during training.

\subsubsection{\textbf{Dataset}} \leavevmode \\
Currently, there are many deepfake detection datasets available, which include both real and manipulated videos and images. To name a few, we have FaceForensics++\cite{b15}, UADFV, Google DFD, CelebDF\cite{b13} and the DFDC dataset\cite{b14}. The source of videos for all datasets is different, and the fake videos are generated from the real frames using various deepfake generation models.
\begin{figure}
\begin{center}
\includegraphics[scale = 0.70]{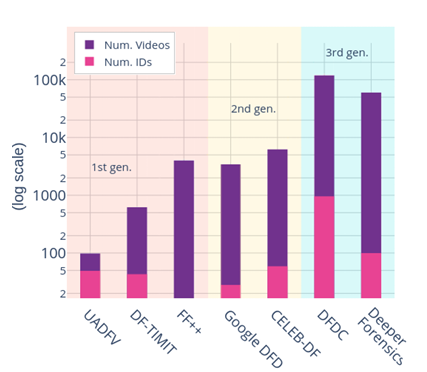}
\end{center}
\caption{Comparison of current deepfakes datasets \cite{imagedyn}}
\label{fig3}
\end{figure}
DFDC is currently the largest available deepfakes dataset (Fig.~\ref{fig3}) with the most number of videos and faces. So we trained the model using DFDC.
The dataset does not ship with the segmentation masks that we need for training the segmentation branch of the U-YNet model. To create the mask, we find the absolute difference between the real frame and the fake frame, which encodes one for a manipulated pixel and a zero for the real one. \\
\textbf{Data Leakage challenge:} Initially, while training we were getting 99\% accuracy on the DFDC dataset, which meant that the model was overfitting. After insinuating on this problem, we found that DFDC has 1:124 videos per subject. So, while randomly splitting the dataset into testing, training, and validation datasets, the training dataset had examples of all the faces in the dataset. So the model was basically learning the faces and not classifying the deepfake features, giving high accuracy on the test dataset. This problem is called data leakage. \\
\textbf{Solution:}
We solved this by clustering similar faces together and then segregating the dataset such that a face in the training dataset does not repeat in the testing dataset. This prevented overfitting, and the model focused on learning features rather than faces.

\subsubsection{\textbf{Loss Function and Training}} \leavevmode \\
For the segmentation task we used Cross Entropy Loss and for the classification task we used Binary Cross Entropy Loss (BCE). \\
Cross-entropy is used to measure the difference between two probability distributions, and it is used for both classification and segmentation tasks. We use the average of both losses as the final loss. \\
The model was trained using the final loss. Finally, Adam Optimiser was used for converging the model. The model was trained for 40 epochs on the created DFDC dataset on NVIDIA DGX-1.

\subsection{Hardware Implementation} 

The hardware setup of Deepfake CLI consists of Dell R740 Server which is a two socket machine with two Intel Xeon processor installed and paired with 32 GB DDR4 RAM running Ubuntu Server 20.04 LTS, kernel version 5.4.0. The server was installed with VCK5000\footnotemark[4]\footnotetext[4]{Read more: https://www.xilinx.com/products/boards-and-kits/vck5000.html}, an FPGA development card built on XILINX 7 nm Versal ACAP\footnotemark[5]\footnotetext[5]{ACAP stands for Adaptive Compute Acceleration Platform. Read more: https://www.xilinx.com/products/silicon-devices/acap/versal.html} Architecture which is ideal for acceleration of AI models for cloud and edge application.
\begin{figure}[ht]
    \centering
    \begin{subfigure}[t]{0.49\textwidth}
        \includegraphics[scale=0.3]{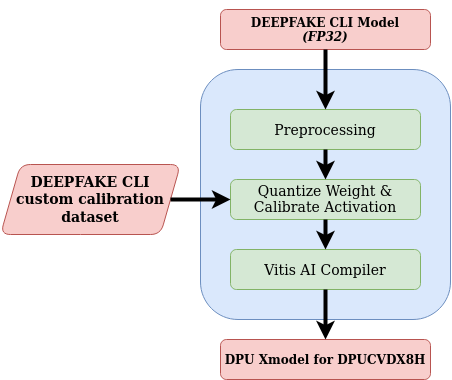}
        \caption{Vitis Flow}
        \label{fig:4}
    \end{subfigure}
    \begin{subfigure}[t]{0.49\textwidth}
        \vspace{-3.1cm}
        \includegraphics[scale=0.3]{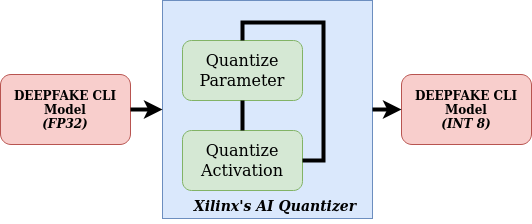}
        \vspace{0.68cm}
        \caption{Step 1: Quantization}
        \label{fig:5}
    \end{subfigure}
    \vspace{6pt}
    \caption{Vitis Flow \& Step 1: Quantization}
\end{figure}\\

In the previous chapter, we have described the changes that had to be done in the dataset to overcome the challenges. For efficient usage of FPGA resources, it is crucial to train a hardware centric model with ideal weights for the model. Since earlier the model was in FP32, and it is not optimal to run this on the FPGA \cite{8bit}, we used the Vitis-AI toolchain to convert the FP32 model to INT8. The entire vitis flow is show in Fig.~\ref{fig:4} which consists of three steps which are Quantization, Compilation, and Inference explained as follows:

\subsubsection{\textbf{Quantization}} \leavevmode \\
Neural nets after being trained on some dataset store all the weights and biases in the FP32 format in the model. If this model is directly used for inference, it will be compute and memory intensive and might not be able to serve the real-time processing demands.
Since detecting deepfake videos in real-time is a crucial goal of this project, to get better inference speed we have quantized our model to be more hardware-centric using Vitis AI Quantizer. This used the Post Training Quantization method to quantize Deepfake CLI model suitable for our real-time needs. Post Training model is less demanding in terms of runtime, being done by running few inference cycles on a dataset known as the calibration dataset. This calibration dataset can be from a training or testing set, and it has to be around 100-1000 images. As shown in Fig.~\ref{fig:5}, we were able to quantize the Deepfake CLI model from FP32 format to INT8 with Quantize parameters and Quantize activations modules.

\subsubsection{\textbf{Compilation}} \leavevmode \\
Once the model quantization step has been completed which generates a .xmodel file, we are ready for the final steps to deploy our model onto VCK5000. We will need to compile our model weights file by using the compiler provided by Vitis-AI. Vitis-AI provides a compiler tool that acts as an interface for optimization of deep neural networks for specific DPUs\footnotemark[6]. With a range of compiler option present due to different DPUs\footnotemark[6] available by Xilinx, it will transform the neural network to the targeted platform DPU instruction set for efficient optimization.

\footnotetext[6]{DPU stands for Deep Learning Processor Unit. Read more: https://www.xilinx.com/products/intellectual-property/dpu.html}

\begin{figure}[ht]
    \centering
    \begin{subfigure}[t]{0.49\textwidth}
        \vspace{-3.05cm}
        \includegraphics[scale=0.3]{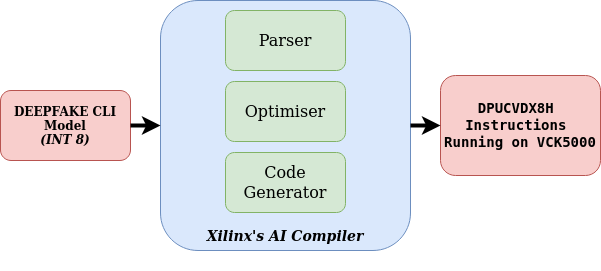}
        \vspace{-0.2cm}
        \caption{Step 2: Compilation}
        \label{fig6a}
    \end{subfigure}
    \begin{subfigure}[t]{0.49\textwidth}
        \hspace*{1.25cm}
        \includegraphics[scale=0.3]{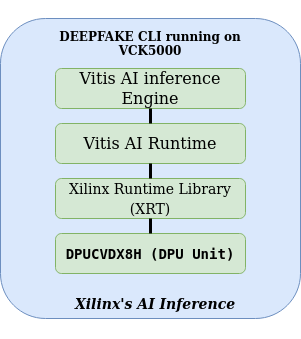}
        \caption{Step 3: Inference}
        \label{fig6b}
    \end{subfigure}
    \vspace{6pt}
    \caption{Hardware Implementation flow}
\end{figure}
Every DPU variant designed by Xilinx to be run on their FPGA board has its own set of optimized instruction to ensure efficient execution of the neural network on hardware. To adhere to this demand, we need to compile our quantized  model for the DPU variant it will run on.
An overview of Vitis AI compiler framework is described in Fig.~\ref{fig6a}. After it has completed topology parsing of the neural network obtained by quantization, the Vitis AI compiler will construct a computational graph as Intermediate Representation. The compiler will also be responsible for further optimization such as fusion of computational nodes which provides higher performance and reduces memory footprint, using the existing parallelism to make sure the instructions are scheduled efficiently.  

\subsubsection{\textbf{Inference}} \leavevmode \\
Now that compilation of our model weights has been completed, we are ready to use it for Inference. We use the Vitis AI Runtime Library(VART) available  in Python and C++. We pre-process the input image by resizing the image to 224×224, and then scale the values from 0-255 to 0-1. The images also have a scaling operation applied to them to convert them to INT8 format as required by the model. We then append the images to a list and pass it to the DPU running on the VCK5000 loaded with our quantized model. \\
VCK5000 takes the images in batches of eight and gives us two outputs, the segmented mask, and a classification label that tells us if the image or video is a deepfake content or not. The segmented output is stored in a .npy format for post-processing, so we could visualize the output in a Jupyter notebook. \\
The classification output is given at the end of the inference code, which tells us how well the model performed on unseen data. A JSON file contains the label for unseen data, we compare it with the model output and print the accuracy of how well it performs. The inference process was explained in the Fig.~\ref{fig6b}.

\section{Results} 
\subsection{Software Results} 
The Deepfake CLI model was evaluated on a testing dataset which had around 2,00,000 frames. These frames have faces that were never seen by the model before. Our main task was classification, hence segmentation results are not presented here. As we can see from the Table~\ref{tab1}, we were able to increase the accuracy of the model by almost 6\% using a multitask learning model.

\begin{table}[htbp]
\begin{center}
\scalebox{1}{
\begin{tabular}{|c|c|}
\hline
\textbf{Model} & \textbf{Accuracy}\\
\hline
 U-YNet model  & 94.12\%\\
\hline
 U-YNet model without segmentation branch   & 88.27\% \\
\hline
\end{tabular}}
\end{center}
\caption{Module Accuracy Comparison}
\label{tab1}
\end{table}

\begin{figure} [ht]
\begin{center}
\includegraphics[scale=0.6]{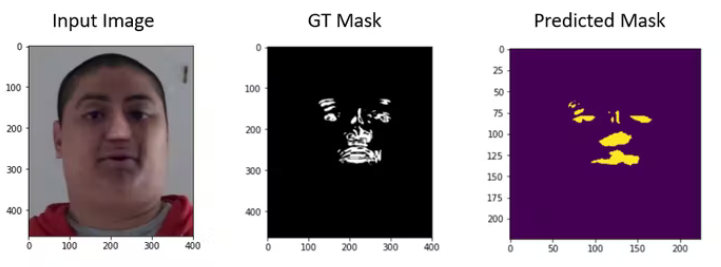}
\caption{(Left to Right) Input Image, Ground Truth, Predicted Mask}
\label{fig7}
\end{center}
\end{figure}

\subsection{Hardware Results} 

\begin{figure} [ht!]
\begin{center}
\includegraphics[scale=0.55]{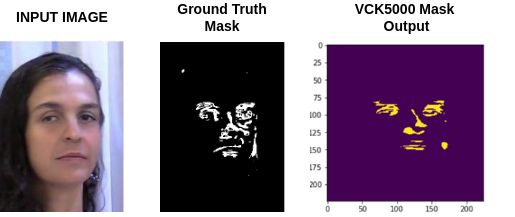}
\caption{(Left to Right) Input Image, Ground Truth, FPGA Output}
\label{fig8}
\end{center}
\end{figure}

The Deepfake CLI model was first quantized, compiled and then inferenced on the VCK5000 FPGA card. We have tested the model on the FPGA with various deepfake videos and images. In the Fig.~\ref{fig7}, the predicted mask by Deepfake CLI model matches with the Ground Truth Mask for the given input image. Now for the hardware results, as shown in Fig.~\ref{fig8} the segmentation output of VCK5000 also matches with the Ground Truth Mask.
Deepfake CLI's main aim was to increase the inference speed of the deepfake detection. Here are the results of the classification branch, shown in the Table~\ref{tab2}.
\begin{table}[htbp]
\begin{center}
\scalebox{1}{
\begin{tabular}{|c|c|}
\hline
\textbf{Parameter} & \textbf{Value}\\
\hline
Total Images & 5001 \\
\hline
Correct Predictions  & 4851 \\
\hline
Wrong Predictions & 350 \\
\hline
Accuracy & 93.001 \% \\
\hline 
\textbf{FPS} & \textbf{316.80} \\
\hline
\end{tabular}}
\end{center}
\caption{Classification Results}
\label{tab2}
\end{table}
We have obtained 316.80 FPS for the classification model of deepfake detection on VCK5000. Because of quantization of the model to INT8, there was around 1\% loss in accuracy, but we got better inference speeds than other inference nodes.
\subsection{Benchmarking Deepfake CLI with other nodes}
We had created a dedicated hardware centric model and inferenced it on VCK5000. We had compared the results of the segmentation model on different state-of-the-art inference nodes available in the market.
\begin{figure}[ht!]
\begin{center}
\includegraphics[scale=0.4]{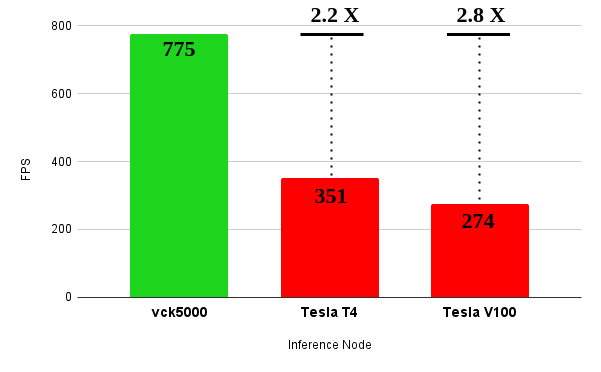}
\end{center}
\caption[FPS vs Inference Nodes]{FPS vs Inference Nodes \footnotemark[7]\footnotemark[8]}
\label{fig9}
\end{figure}
\footnotetext[7]{All nodes were running INT8 Deepfake CLI Unet model, TensorRT SDK was used to quantize INT8 model for Tesla T4 and V100 while for VCK5000 we have used Vitis AI Flow.}
\footnotetext[8]{In this benchmarking test, we have compared the inference speed of simpler Deepfake CLI model which only consists of Unet model (segmentation branch). This was done due to the TensorRT quantization limitation for Tesla T4 and V100 nodes. We have also obtained the complete Y-UNet Inference speed for VCK5000, which was 316.80 FPS (Table~\ref{tab2}). When both branches are included in Deepfake CLI, the layers in the network gets approximately doubled and thus reduces inference speed by half.}
\\ As shown in Fig.~\ref{fig9}, we have obtained 775 FPS for the segmentation model of deepfake detection on VCK5000. For the same model, we had got a much lower FPS on other inference nodes, like Nvidia's Tesla T4 and V100. This result shows that with proper model preprocessing, quantization, and compilation for FPGA architecture will give better inference speeds than its counterpart GPUs.

\section{Conclusion}Because of the rapid increase in computation speeds, deepfake detection will be one of the concerns in the upcoming era. In this paper, we implemented a novel architecture that combined segmentation and classification models, which were trained on DGX A100 and inferred on VCK5000. We used a modified DFDC dataset to reduce data leakage and solve the over fitting problem. The Deepfake CLI software model was designed from a hardware-centric point of view, and it was quantized and compiled for the VCK5000 AI inference versal card. We were able to classify the deepfake images as well as videos on the Accelerated FPGA fabric. \textbf{We have obtained a 120\% gain on the inference speed on the VCK5000 FPGA fabric compared to state-of-the-art Tesla T4 inference node.}

\section{Future Work}
The ever-increasing research in deepfake generation demands better and faster techniques to decode and detect these deepfake images and videos in real-time. The future goals of Deepfake C-L-I are as follows:
\begin{itemize}
   \item Modeling and Training more generalized model for deepfake detection.
   \item Restoring the deepfaked region of the frame.
   \item Adding AWS FPGA support for easy access.
   \item Adding accelerators for pre and post-processing of images and videos for ML model.
\end{itemize}

\section{Acknowledgment}

We are grateful to the Centre of Excellence in Complex and Non-Linear Dynamical Systems, V.J.T.I. (\emph{CoE-CNDS}) for support in the project. Also, we would like to thank the entire hackster.io and AMD-Xilinx Team for shipping us the VCK5000 board, and the technical support to get it up and running.

%
%
%

\end{document}